\documentstyle{article}
\title{Normalized Weyl-type $\star$-product on~K\"{a}hler~manifolds}
\author
{
Takuya MASUDA\footnote{tmasuda@phys.metro-u.ac.jp} \\ \ \\
{\it Department of Physics, Tokyo Metropolitan University}, \\{\it Hachioji, Tokyo, 192-0397, Japan}
}
\begin{document}
\maketitle
\begin{abstract}
We define a normalized Weyl-type $\star$-product on general K\"{a}hler
manifolds. Expanding this product perturbatively we show that the
cumbersome term, which appears in a Berezin-type product, does not appear at
least in the first order of $\hbar$. This means a normalization factor,
 which is introduced by Reshetikhin and Takhtajan for a Berezin-type product, is 
unnecessary for our Weyl-type product at that order.
\end{abstract}
\section{Introduction}

A new kind of mathematics is thought necessary for a
non-perturbative description of the string theory just as Riemannian
geometry is indespensable for the description of the theory of general
relativity. Non-commutative geometry is one of the strong candidate for
it.

We would like to construct a non-commutative manifold with a K\"{a}hler
metric from this perspective. We take deformation quantization approach,
which introduces a non-commutative product into the ring of functions on
a commutative manifold. Kontsevich showed how to construct a
non-commutative product on Poisson manifolds, which includes K\"{a}hler manifolds~\cite{Kontsevich}~\cite{Felder}, but the effect of
a metric is implicit in their construction. Dependence of a
non-commutativity on a metric is clear in our construction. We give an
outline of our construction in the rest part of this section.

Among other approaches we focus our attention on two different types of
non-commutative products, the Weyl-type and the Berezin-type. The
ordinary Moyal product of functions $f_1$ and $f_2$ on ${\rm
C}^1$~\cite{Moyal}~\cite{Groenewold} can be written in an integral representation,
as
%\newpage
\newcommand{\vbar}{\overline{v}}
\newcommand{\wbar}{\overline{w}}
\newcommand{\zbar}{\overline{z}}
\[
\left(f_1 *_M f_2\right)(z,\overline{z})=\int_{{\rm
C}^1}f_1(w,\overline{v})f_2(v,\overline{w})\frac{e^{\left(z\overline{v}+v\overline{z}-z\overline{z}-v\overline{v}\right)/\hbar}}{e^{\left(z\overline{w}+w\overline{z}-z\overline{z}-w\overline{w}\right)/\hbar}}\frac{dvd\vbar}{2\pi i\hbar}\frac{dwd\wbar}{2\pi i\hbar},
\]
which reduces to
\[
 f_1f_2+\frac{\hbar}{2\sqrt{-1}}\left\{f_1,f_2\right\}_P+O\left(\hbar^2\right)
\]
in the small $\hbar$ limit and satisfies associativity. This approach of
quantization is applicable only to a flat manifold.

There are papers on Berezin-type star-product on K\"{a}hler manifolds~\cite{Cahen}~\cite{Karabegov}.
The Berezin star-product is defined on certain K\"{a}hler
manifolds by
\begin{equation}
\left(f_1\bullet f_2\right)(z,\overline{z})=\int_{{\rm
C}^n}f_1(z,\overline{v})f_2(v,\overline{z})e^{\left(\Phi(z,\overline{v})+\Phi(v,\overline{z})-\Phi(z,\overline{z})-\Phi(v,\overline{v})\right)/\hbar}d\mu_\hbar(v,\overline{v})\label{NNBerezin}
\end{equation}
where $\Phi$ is a K\"{a}hler potential of the manifold.
This product also satisfies the associativity. The reduction to the
Poisson bracket is achieved, however, only in the difference of two
terms:
\[
 f_1\bullet f_2-f_2\bullet f_1=\frac{\hbar}{2\sqrt{-1}}\left\{f_1,f_2\right\}_P+O\left(\hbar^2\right).
\]

Reshetikhin and Takhtajan showed (\ref{NNBerezin}) does not necessarily
satisfy $f\bullet 1=1\bullet f=f$ for the general K\"{a}hler potentials
and defined a product which satisfies the above equality.

It was shown in~\cite{Wakatsuki} that there exists another possible
approach

\begin{equation}
\left(f_1\odot f_2\right)(z,\overline{z})=\int_{{\rm C}^n}f_1(w,\overline{v})f_2(v,\overline{w})\frac{e^{\left(\Phi(z,\overline{v})+\Phi(v,\overline{z})-\Phi(z,\overline{z})-\Phi(v,\overline{v})\right)/\hbar}}{e^{\left(\Phi(z,\overline{w})+\Phi(w,\overline{z})-\Phi(z,\overline{z})-\Phi(w,\overline{w})\right)/\hbar}}d\mu_\hbar(v,\overline{v})d\mu_\hbar(w,\overline{w})\label{NNSP}
\end{equation}
which interpolates between the Weyl-type and Berezin-type
star-products. In the flat space this is the same as the Moyal
product. The associativity, however, does not hold by itself, but is
fulfilled in the functional integral limit of its multiple products.

Ref.~\cite{Wakatsuki}
is insufficient in the sense that there is no consideration to the general K\"{a}hler manifolds for which
$A=\frac{1}{2}\sum_{i,\overline{j}=1}^nh^{i\overline{j}}\partial_i\overline{\partial}_j\log\det 
H$ is not 0, where the matrix $H$ is a metric.

We generalize the product defined in~\cite{Wakatsuki} so that it is
applicable to general K\"{a}hler manifolds. Moreover we expand the
product perturbatively and show that it really has a property
characteristic of
a Weyl-type product and that the normalization factor necesarry for a
Berezin-type product introduced by Reshetikhin and Takhtajan is unnecessary for a Weyl-type product at least in
the first order of $\hbar$.

\section{Construction}

Reshetikhin and Takhtajan showed the
Berezin-type product (\ref{NNBerezin})
does not necessarily satisfy $f\bullet 1=1\bullet
f=f$ for the general K\"{a}hler potentials and defined, as a product
which satisfies the above equality, a normalized star-product
\begin{equation}
 e_\hbar^{-1}(z,\overline{z})\left((f_1e_\hbar)\bullet(f_2e_\hbar)\right)(z,\overline{z})\label{ehbar},
\end{equation}
where $e_\hbar$ is a normalization factor defined as $f\bullet e_\hbar=e_\hbar\bullet
f=f$~\cite{Takhtajan}.

As you can see, (\ref{NNSP}) consists of two kinds of
Berezin-type products, an ordinary Berezin-type product
\begin{eqnarray*}
\left(f_1\bullet f_2\right)(z,\overline{z})&=&\int_{{\rm C}^n}f_1(z,\overline{v})f_2(v,\overline{z})e^{\left(\Phi(z,\overline{v})+\Phi(v,\overline{z})-\Phi(z,\overline{z})-\Phi(v,\overline{v})\right)/\hbar}d\mu_\hbar(v,\overline{v})
\end{eqnarray*}
and a new kind of Berezin-type product
\begin{eqnarray*}
\left(f_1\circ f_2\right)(z,\overline{z})&:=&\int_{{\rm C}^n}f_1(v,\overline{z})f_2(z,\overline{v})e^{-\left(\Phi(z,\overline{v})+\Phi(v,\overline{z})-\Phi(z,\overline{z})-\Phi(v,\overline{v})\right)/\hbar}d\mu_\hbar(v,\overline{v}).
\end{eqnarray*}
\newcommand{\ehat}{\hat{e}}
A normalization factor is necessary for each kind of Berezin-type
product, so a normalization factor is also necessary for a Weyl-type
product.

We define a normalization factor $\ehat_\hbar$ for a new kind of
Berezin-type product $f_1\circ f_2$ so that $f\circ \hat{e}_\hbar=\hat{e}_\hbar\circ f=f$. 
\[
 e_\hbar=1-\hbar A+O\left(\hbar^2\right)
\]
whereas
\[
 \ehat_\hbar=1+\hbar A+O\left(\hbar^2\right).
\]

We define a normalized star-product for $f_1\circ f_2$ just as (\ref{ehbar})
for $f_1\bullet f_2$ as
\[
 \ehat_\hbar^{-1}(z,\overline{z})\left((f_1\ehat_\hbar)\circ(f_2\ehat_\hbar)\right)(z,\overline{z}).
\]

With two kinds of normalization factors $e_\hbar$ and $\ehat_\hbar$ we
define a normalized Weyl-type prodct $f_1* f_2$ as
\begin{eqnarray}
&&\left(f_1* f_2\right)(z,\overline{z})\nonumber\\
&:=&e_\hbar^{-1}(z,\overline{z})\left((e_\hbar f_1 \hat{e}_\hbar)\odot (e_\hbar f_2 \hat{e}_\hbar)\right)(z,\overline{z})\hat{e}_\hbar^{-1}(z,\overline{z})\label{NSP}\\
&=&\int_{{\rm C}^n}f_1(w,\overline{v})f_2(v,\overline{w})\frac{e_\hbar(w,\overline{v})e_\hbar(v,\overline{w})}{e_\hbar(z,\overline{z})}\frac{\hat{e}_\hbar(w,\overline{v})\hat{e}_\hbar(v,\overline{w})}{\hat{e}_\hbar(z,\overline{z})}\frac{e^{\Phi(z',\overline{v})+\Phi(v,\overline{z})-\Phi(z',\overline{z})-\Phi(v,\overline{v})}}{e^{\Phi(z,\overline{w})+\Phi(w,\overline{z}')-\Phi(z,\overline{z}')-\Phi(w,\overline{w})}}d\mu_\hbar(v,\overline{v})d\mu_\hbar(w,\overline{w})\nonumber
\end{eqnarray}
This product satisfies
\[
 f* 1=1* f=f
\]
for general K\"{a}hler manifolds. It is clear from (\ref{NSP}) generally
\mbox{$f_1\odot f_2\ne f_1* f_2$} unless $\ehat_\hbar=e_\hbar^{-1}$.

As non-normalized star-product $f_1\odot f_2$, normalized
star-product $f_1* f_2$ does not satisfy the associativity in
this form, but the transition to the functional integral version goes as follows.

\

Multi-products from both sides $\left(f^{(0)}* \left(\cdots* \left(f^{(N-2)}*
\left(f^{(N-1)}* f^{(N)}\right)\right)\cdots\right)\right)$ and 
$\left(\left(\cdots\left(\left(f^{(0)}* f^{(1)}\right)*
f^{(2)}\right)* \cdots\right)* f^{(N)}\right)$
are respectively
\begin{eqnarray*}
&&\left(f^{(0)}* \left(\cdots* \left(f^{(N-2)}* \left(f^{(N-1)}* f^{(N)}\right)\right)\cdots\right)\right)\left(z,\zbar\right)\\
&=&\int_{{\rm C}^n}\left(\prod_{j=1}^Nd\mu_\hbar\left(z^{(j-1)},\vbar^{(j)}\right)d\mu_\hbar\left(v^{(j)},\zbar^{(j-1)}\right)\frac{e^{\phi\left(v^{(j-1)},\vbar^{(j-1)};v^{(j)},\zbar^{(j-1)}\right)}}{e^{\phi\left(v^{(j-1)},\vbar^{(j-1)};z^{(j-1)},\vbar^{(j)}\right)}}e_\hbar\left(z^{(j-1)},\zbar^{(j-1)}\right)\hat{e}_\hbar\left(z^{(j-1)},\zbar^{(j-1)}\right)\right)\\
&\times&\frac{e_\hbar\left(z^{(N)},\zbar^{(N)}\right)}{e_\hbar\left(z,\zbar\right)}\frac{\hat{e}_\hbar\left(z^{(N)},\zbar^{(N)}\right)}{\hat{e}_\hbar\left(z,\zbar\right)}\left(\prod_{j=0}^Nf^{(j)}\left(z^{(j)},\zbar^{(j)}\right)\right)\qquad\left(v^{(N)}=z^{(N)},\quad v^{(0)}=z\right),
\end{eqnarray*}
\begin{eqnarray*}
&&\left(\left(\cdots\left(\left(f^{(0)}* f^{(1)}\right)* f^{(2)}\right)* \cdots\right)* f^{(N)}\right)\left(z,\zbar\right)\\
&=&\int_{{\rm C}^n}\left(\prod_{j=1}^N d\mu_\hbar\left(z^{(j)},\vbar^{(j-1)}\right)d\mu_\hbar\left(v^{(j-1)},\zbar^{(j)}\right)\frac{e^{\phi\left(v^{(j)},\vbar^{(j)};z^{(j)},\vbar^{(j-1)}\right)}}{e^{\phi\left(v^{(j)},\vbar^{(j)};v^{(j-1)},\zbar^{(j)}\right)}}e_\hbar\left(z^{(j)},\zbar^{(j)}\right)\hat{e}_\hbar\left(z^{(j)},\zbar^{(j)}\right)\right)\\
&\times&\frac{e_\hbar\left(z^{(0)},\zbar^{(0)}\right)}{e_\hbar\left(z,\zbar\right)}\frac{\hat{e}_\hbar\left(z^{(0)},\zbar^{(0)}\right)}{\hat{e}_\hbar\left(z,\zbar\right)}\left(\prod_{j=0}^Nf^{(j)}\left(z^{(j)},\zbar^{(j)}\right)\right)\qquad\left(v^{(0)}=z^{(0)},\quad v^{(N)}=z\right).
\end{eqnarray*}
Therefore they have the same functional integral limit,
\begin{eqnarray*}
&&\int{\cal D}\mu\left(z,\overline{v}\right){\cal
D}\mu\left(v,\overline{z}\right)\\
&\times&\exp\left[\int d\tau\left\{
{\frac{\partial \vbar}{\partial \tau}\left(\frac{\partial\Phi\left(z,\vbar\right)}{\partial\vbar}-\frac{\partial
\Phi\left(v,\vbar\right)}{\partial \vbar}\right)}
-\frac{\partial v}{\partial \tau}\left(\frac{\partial
\Phi\left(v,\zbar\right)}{\partial
v}-\frac{\partial\Phi\left(v,\vbar\right)}{\partial v}\right)+\log
e_\hbar\left(z,\zbar\right)+\log \hat{e}_\hbar\left(z,\zbar\right)+\log f\left(z,\overline{z}\right)\right\}\right].
\end{eqnarray*}
as in~\cite{Wakatsuki}, where
$f(z,\overline{z})$ is defined as
\[
 \int d\tau\ \log
f\left(z,\overline{z}\right)=\lim_{N\to\infty}\sum_{j=0}^N\frac{1}{N}\log
f^{(j)}\left(z^{\left(j\right)},\overline{z}^{\left(j\right)}\right)\qquad 
\left(z^{(j)}-z^{(j-1)}=\frac{1}{N}\right).
\]
>From the same discussion as in~\cite{Wakatsuki}, we define a normalized
associative Weyl-type product as
\begin{eqnarray*}
&&f_1\star f_2\\
&=&\int{\cal D}\mu\left(z,\overline{v}\right){\cal D}\mu\left(v,\overline{z}\right)\\
&\times&f_1\left(z\left(\tau_1\right),\overline{z}\left(\tau_1\right)\right)f_2\left(z\left(\tau_1\right),\overline{z}\left(\tau_2\right)\right)\exp\left[\int d\tau\left\{\right.\right.\\
&&\left.\left.
{\frac{\partial \vbar}{\partial \tau}\left(\frac{\partial\Phi\left(z,\vbar\right)}{\partial\vbar}-\frac{\partial
\Phi\left(v,\vbar\right)}{\partial \vbar}\right)}
-\frac{\partial v}{\partial \tau}\left(\frac{\partial
\Phi\left(v,\zbar\right)}{\partial
v}-\frac{\partial\Phi\left(v,\vbar\right)}{\partial v}\right)+\log
e_\hbar\left(z,\zbar\right)+\log \hat{e}_\hbar\left(z,\zbar\right)\right\}\right],
\end{eqnarray*}
where $\tau_1$ and $\tau_2$ are fixed points.

\section{Perturbation}

We expand $f_1\odot f_2$ perturbatively in small $\hbar$. The perturbative expansion of
one of the Berezin-type products is given by~\cite{Takhtajan} :
\newcommand{\partialbar}{\overline{\partial}}
\newcommand{\ybar}{\overline{y}}
\newcommand{\rec}[1]{\frac{1}{#1}}
\begin{eqnarray}
&&f_1\bullet
f_2\nonumber\nonumber\\
&=&\pi^{-n}\det H\int_{{\rm C}^n}e^{-\left(Hy,y\right)}\prod_{i=1}^n\frac{\left|dy^i\wedge d\overline{y}^i\right|}{2}\left[\frac{}{}f_1f_2+\epsilon^2\left\{y^i\ybar^j\left(\partialbar_j f_1\right)\left(\partial_i f_2\right)\frac{}{}\right.\right.\nonumber\\
 &+&\left.\left.f_1f_2\left(-\rec{4}y^iy^j\ybar^k\ybar^\ell\partial_i\partialbar_k \Phi_{j\overline{\ell}}+y^i\ybar^j\frac{\left(\partial_i\partialbar_j\det H\right)\det H-\left(\partial_i\det H\right)\left(\partialbar_j\det H\right)}{\left(\det H\right)^2}\right)\right.\right.\nonumber\\
&+&y^i\ybar^j\left(f_2\partialbar_j f_1\frac{\partial_i\det H}{\det H}+f_1\partial_i f_2\frac{\partialbar_j\det H}{\det H}\right)\nonumber\\
 &+&f_1f_2\left(\rec{4}y^iy^jy^k\ybar^\ell\ybar^m\ybar^n\left(\partial_i \Phi_{j\overline{k}}\right)\left(\partialbar_\ell \Phi_{m\overline{n}}\right)+y^i\ybar^j\frac{\left(\partial_i\det H\right)\left(\partialbar_j\det H\right)}{\left(\det H\right)^2}\right.\nonumber\\
 &-&\left.\left.\left.\rec{2}y^iy^j\ybar^k\ybar^\ell\left\{
\frac{\left(\partial_i \Phi_{j\overline{k}}\right)\left(\partialbar_\ell \det H\right)}{\det H}
+\frac{\left(\partial_i \det H\right)\left(\partialbar_k \Phi_{j\overline{\ell}}\right)}{\det H}
\right\}\right)\right\}+O\left(\epsilon^3\right)\right]\nonumber\\
&=&f_1f_2+\hbar\left(Af_1f_2+\sum_{i\overline{j}}h^{i\overline{j}}\partialbar_j f_1\partial_if_2\right)+O\left(\hbar^2\right),\quad\left(\hbar=\epsilon^2\right)\label{bullet}
\end{eqnarray}

We apply this method of perturbation to the new Berezin-type product,
$f_1\circ f_2$ we have defined above :
\begin{eqnarray}
&&f_1\circ
f_2\nonumber\nonumber\\
&=&\pi^{-n}\det H\int_{{\rm C}^n}e^{\left(Hy,y\right)}\prod_{i=1}^n\frac{\left|dy^i\wedge d\overline{y}^i\right|}{2}\left[\frac{}{}f_1f_2+\epsilon^2\left\{y^i\ybar^j\left(\partial_i f_1\right)\left(\partialbar_j f_2\right)\frac{}{}\right.\right.\nonumber\\
 &+&\left.\left.f_1f_2\left(\rec{4}y^iy^j\ybar^k\ybar^\ell\partial_i\partialbar_k \Phi_{j\overline{\ell}}+y^i\ybar^j\frac{\left(\partial_i\partialbar_j\det H\right)\det H-\left(\partial_i\det H\right)\left(\partialbar_j\det H\right)}{\left(\det H\right)^2}\right)\right.\right.\nonumber\\
&+&y^i\ybar^j\left(f_1\partialbar_j f_2\frac{\partial_i\det H}{\det H}+f_2\partial_i f_1\frac{\partialbar_j\det H}{\det H}\right)\nonumber\\
 &+&f_1f_2\left(\rec{4}y^iy^jy^k\ybar^\ell\ybar^m\ybar^n\left(\partial_i \Phi_{j\overline{k}}\right)\left(\partialbar_\ell \Phi_{m\overline{n}}\right)+y^i\ybar^j\frac{\left(\partial_i\det H\right)\left(\partialbar_j\det H\right)}{\left(\det H\right)^2}\right.\nonumber\\
 &+&\left.\left.\left.\rec{2}y^iy^j\ybar^k\ybar^\ell\left\{
\frac{\left(\partial_i \Phi_{j\overline{k}}\right)\left(\partialbar_\ell \det H\right)}{\det H}
+\frac{\left(\partial_i \det H\right)\left(\partialbar_k \Phi_{j\overline{\ell}}\right)}{\det H}
\right\}\right)\right\}+O\left(\epsilon^3\right)\right]\nonumber\\
&=&f_1f_2-\hbar\left(Af_1f_2+\sum_{i\overline{j}}h^{i\overline{j}}\partial_if_1\partialbar_jf_2\right)+O\left(\hbar^2\right)\quad\left(\hbar=\epsilon^2\right)\label{circ}.
\end{eqnarray}
As a result we get
\begin{equation}
f_1\odot f_2=f_1f_2+\hbar\sum_{i,j=1}^nh^{i\overline{j}}\left(\partialbar_jf_1\partial_if_2-\partial_if_1\partialbar_jf_2\right)+O\left(\hbar^2\right).
\end{equation}

A Poisson bracket appears in the first order of $\hbar$, which means the
product $f_1\odot f_2$ is really a Weyl-type product and which was not
shown in~\cite{Wakatsuki}.

What is surprising is the disapperance of a
term
$A=\frac{1}{2}h^{i\overline{j}}\partial_i\partialbar_j\log\det H$ in
the first order. Compare this result with the expansion of $f_1\bullet
f_2$ in~\cite{Takhtajan}, namely (\ref{bullet}), and of $f_1\circ f_2$.
The cumbersome term A, which appears in a Berezin-type product, disappears
in a Weyl-type product at least in the order of $\hbar$. This means, therefore, the
normalization factor is unnecessary for the Weyl-type product at least
in the first order of $\hbar$.

\

In conclusion we have found that the Weyl-type product defined
in~\cite{Wakatsuki} can be decomposed into two kinds of Berezin-type
products and introduce a normalization factor into a Weyl-type product
since a normalization factor is necessary for each kind of Berezin-type
product. Note, in
general, we must use $f_1* f_2\left(\ne f_1\odot f_2\ {\rm unless}\ \ehat_\hbar=e_\hbar^{-1}\right)$.

Moreover we perform a perturbative expansion of non-normalized
Weyl-type product. In the first order of $\hbar$ a Poisson bracket
appears and $A$ does not appear. Therefore the normalization factor
which is necessary for a Berezin-type product is unnecessary for a
Weyl-type product at least in the first order of $\hbar$. The
normalization factor is unnecessary for Weyl-type product
non-perturbatively if $\ehat_\hbar=e_\hbar^{-1}$ is shown.

\

\noindent
{\large {\bf Acknowledgement}}

\

I would like to thank Dr. Saito for useful discussions and his correction and Dr. Iso for his introduction
and explanation of~\cite{Takhtajan} and Mr. Wakatsuki for his
cooperation in reading~\cite{Takhtajan}.

\end{document}